# Giant and tunable excitonic optical anisotropy in single-crystal CsPbX₃ halide perovskites

*G.A. Ermolaev[1], A.P. Pushkarev[2], A.Yu. Zhizhchenko[3,4], A.A. Kuchmizhak[3,4], I.V. Iorsh[2], I. Kruglov[5,6], A. Mazitov[5,6], A. Ishteev[7,8], K. Konstantinova[7], D. Saranin[7], A.S. Slavich[5], D. Stosic[5], E. Zhukova[5], G. Tselikov[1], Aldo Di Carlo[7,9], A.V. Arsenin[1], K.S. Novoselov[10,11,12], S.V. Makarov[2], and V.S. Volkov[1]*

[1]Xpanceo, Dubai Investment Park First, Dubai, UAE.

[2]ITMO University, School of Physics and Engineering, St. Petersburg, 197101, Russia

[3]Far Eastern Federal University, Vladivostok 690091, Russia

[4]Institute of Automation and Control Processes, Far Eastern Branch, Russian Academy of Science, Vladivostok 690041, Russia

[5]Center for Photonics and 2D Materials, Moscow Institute of Physics and Technology, Dolgoprudny 141700, Russia.

[6]Dukhov Research Institute of Automatics (VNIIA), Moscow 127055, Russia.

[7]LASE – Laboratory of Advanced Solar Energy, NUST MISiS, 119049 Moscow, Russia

[8]N.N. Semenov Federal Research Center for Chemical Physics, Russian Academy of Sciences 119991, Moscow, Russia

[9]CHOSE – Centre of Hybrid and Organic Solar Energy, Department of Electronics Engineering, Rome, Italy

[10]National Graphene Institute (NGI), University of Manchester, Manchester M13 9PL, UK.

[11]Institute for Functional Intelligent Materials, National University of Singapore, Singapore.

[12]Chongqing 2D Materials Institute, Chongqing 400714, China.

e-mail: volkov.vs@mipt.ru

## ABSTRACT

During the last years, giant optical anisotropy demonstrated its paramount importance for light manipulation which resulted in numerous applications ranging from subdiffraction light guiding to switchable nanolasers. In spite of recent advances in the field, achieving continuous tunability of optical anisotropy remains an outstanding challenge. Here, we present a solution to the problem through chemical alteration of the ratio of halogen atoms (X = Br or Cl) in single-crystal $CsPbX_3$ halide perovskites. It turns out that the anisotropy originates from an excitonic resonance in the perovskite, which spectral position and strength are determined by the halogens composition. As a result, we manage to continually modify the optical anisotropy by 0.14. We also discover that the halide perovskite can demonstrate optical anisotropy up to 0.6 in the visible range – the largest value among non-van der Waals materials. Moreover, our results reveal that this anisotropy could be in-plane and out-of-plane, depending on perovskite shape – rectangular and square. Hence, it can serve as an additional degree of freedom for anisotropy manipulation. As a practical demonstration, we created perovskite anisotropic nanowaveguides and show a significant impact of anisotropy on high-order guiding modes. These findings pave the way for halide perovskites as a next-generation platform for tunable anisotropic photonics.

## INTRODUCTION

Optical anisotropy is one of the most useful and thought after effects for modern nanophotonics.[1] Aside from additional degree of freedom,[2] anisotropy enables novel physical effects and applications: extreme skin-depth guiding,[3] ultrastrong coupling,[4] polariton canalization,[5,6] achromatic waveplates,[7] switching nanolasers,[8] exciton-polariton transport,[3,9] lossless light confinement,[10] exceptional coupling,[11] topological singularities,[12] ghost[13] and shear[14] polaritons to name just a few. These breakthroughs establish a quest for anisotropic materials with giant and tunable birefringence $\Delta n$, which describes the speed of light difference between orthogonal directions.[15] The latest studies[3,16] propose the usage of van der Waals layered materials to tackle this problem since weak binding between atoms in one of the directions naturally lead to high birefringence. Additionally, layered structures are susceptible to atom intercalation, which is an efficient route for anisotropy manipulation.[17,18] Such methods of controlling anisotropy, however, do not allow for continuous adjustment, resulting in a discrete number of material phases. Furthermore, the intercalation approach has been implemented only for mid-infrared spectral range so far,[17,18] while the majority of photonic devices operate at visible frequencies.[12,19,20] As a result, tunable anisotropic optical response is in high demand.

In turn, halide perovskites is a family of materials where variation of anions in the composition can dramatically change band gap over the whole visible range.[21] In particular, the optical properties of perovskite composition $CsPbBr_{3-x}Cl_x$ can be easily tuned by HBr or HCl acids exposure even in gas phase.[21,22] More importantly, halide perovskites, for instance $CsPbBr_3$, exhibit orthorhombic crystal structure (**Figure 1a**) at room temperature,[8,23,24] thus naturally having anisotropic optical response. Surprisingly, only recent research[7,8,25] paid attention to anisotropy, while many others[26,27] continue to use the isotropic treatment. This inconsistency between works leads to a question: why anisotropy manifest itself only in limited cases. For this reason, accurate characterization of anisotropic dielectric tensor of halide perovskites is of paramount importance for advanced optics and perovskite applications.

In this work, we resolve this challenge and demonstrate a broadband modification of the anisotropic optical properties of $CsPbBr_{3-x}Cl_x$ through the variation of chemical atomic composition. We found that halide perovskite anisotropy is comparable to classical birefringent materials such as rutile and calcite at near-infrared wavelengths. Meanwhile, at exciton resonance, anisotropy reaches much higher values, exceeding all non-van der Waals materials. Additionally, our study shows that shape can determine the direction of optical axis and, therefore, anisotropic tensor. As a practical demonstration, we present anisotropic waveguides based on $CsPbBr_3$. Remarkably, our results are applicable at all scales: nano, micro, and macro, unlike layered materials, which to date demonstrate giant anisotropy only at the microscale. Moreover, halide perovskites possess countless benefits: low-cost fabrication,[28,29] high quantum yield of luminescence[30] and optical contrast,[31] strong excitonic response,[32,33] and wide composition diversity.[34,35] Therefore, the proposed tunable optical anisotropy in single-crystal $CsPbX_3$ offers an unprecedented platform for current and future perovskite nanophotonics and optoelectronics.

## RESULTS

**Optical anisotropy in halide perovskites microcrystals**

Orthorhombic structure ($a$ = 8.2563 Å, $b$ = 8.2012 Å, $c$ = 11.7308 Å) of halide perovskites, presented in **Figure 1a**, essentially results in different optical responses along orthogonal directions. These responses can be described by diagonal dielectric tensor diag($n_a, n_b, n_c$), where $n_a$, $n_b$, and $n_c$ are complex refractive indices along corresponding crystallographic basis ($a,b,c$). Provided that $a \approx b$ within 0.7% accuracy and, hence, $n_a \approx n_b = n_{ab}$, halide perovskites are basically uniaxial crystals with dielectric tensor diag($n_{ab}, n_{ab}, n_c$). Our polarized microtransmittance measurements in Supplementary Note 1 support this conclusion. To our surprise, polarized microtransmittance also showed that rectangular-shaped microplates have in-plane anisotropy, in contrast to square-shaped with out-of-plane anisotropy. This observation explains the inconsistency between anisotropic and isotropic responses of halide perovskite in recent works.[7,8,25–27]

In order to manipulate the crystal structure we placed $CsPbBr_3$ microplates in HCl atmosphere (**Supplementary Note 2**). HCl reacts with the crystal by substituting Br atoms with Cl atoms, resulting in an intermediate state $CsPbBr_{3-x}Cl_x$, illustrated in **Figure 1b**. In this way, together with the crystal structure, we also control the whole dielectric tensor diag($n_{ab}, n_{ab}, n_c$), described by two critical parameters: refractive index $n_{ab}$ and birefringence $\Delta n = n_{ab} - n_c$. To investigate this effect, we leveraged imaging spectroscopic ellipsometry because it allows us detection of the optical signal even from microplates, **Figure 1c**. In order to boost the signal we chose thick microplates (**Figure 1d**), which demonstrate Fabry-Perot resonances. They enhance the ellipsometer sensitivity to material anisotropy and produce characteristic asymmetric peaks in spectra.[3] Besides, we focused on microplates with out-of-plane anisotropy, which allow standard ellipsometry characterization. During the imaging ellipsometry, we also measure the maps of ellipsometric parameters $\Psi$ (amplitude) and $\Delta$ (phase) (**Figure 1e-f**) from which we can deduce the local response of microplates. As expected, the representative spectra in **Figure 1g** manifests numerous asymmetric oscillations in perfect agreement with anisotropic optical model, shown in **Figure 1h**. At the same time, the isotropic optical model, plotted in **Figure 1i**, cannot describe the experimental data even qualitatively (**Figure 1g**). This demonstrates that halide perovskites indeed have high anisotropy in accordance with their orthorhombic structure.

For ellipsometry analysis, we note high quality of Fabry-Perot resonances in **Figure 1g**, indicating zero extinction coefficient $k$ in the transparency region. At the same time, such resonances are strongly affected by the pronounced excitonic transition. By analogy to transition metal dichalcogenides,[3] in order to describe the dielectric function of halide perovskites we used the Tauc-Lorentz oscillator[36] and Cauchy model[3] for crystallographic *ab*-plane and *c*-axis, respectively (**Methods**). In particular, this description gives ellipsometry spectra in **Figure 1h** close to the actual data in **Figure 1g**. For additional verification, we also recorded the microtransmittance spectra, which coincides with the transfer matrix calculations, presented in **Supplementary Note 3**. Consequently, Tauc-Lorentz and Cauchy models accurately characterize anisotropic optical response of halide perovskites.

Based on this description, we determined the modification of halide perovskites optical properties by HCl reaction. For example, **Figure 1j** displays the results for $CsPbBr_3$ and $CsPbCl_3$, which defines the variation range of our tuning approach. Apart from absolute value change, it also controls the transparency region and can be used in switching photonic elements. **Figures 1k-l** demonstrate monotonous refractive index and birefringence change with respect to stoichiometry number x of Cl atoms in $CsPbBr_{3-x}Cl_x$. Interestingly, we can chemically tune perovskites' refractive index by 0.27 and anisotropy by 0.14. For instance, it allows us to cover anisotropic properties from $InGaS_3$[37] to calcite,[38] as seen from **Figure 1l**. At exciton resonance, however, anisotropy dramatically increases and even exceeds rutile birefringence (**Figure 1l**). Hence, this giant anisotropic response originates from exciton resonance. It enables us to theoretically model optical properties of halide perovskites (see dashed lines in **Figure 1k** and **Supplementary Note 4**). Besides, our density functional theory calculations yields similar results for optical properties (**Supplementary Note 5**) Thus, our tuning of anisotropic dielectric tensor of halide perovskites relies on exciton shift of $CsPbBr_{3-x}Cl_x$ structure by altering the ratio of Br and Cl atoms.

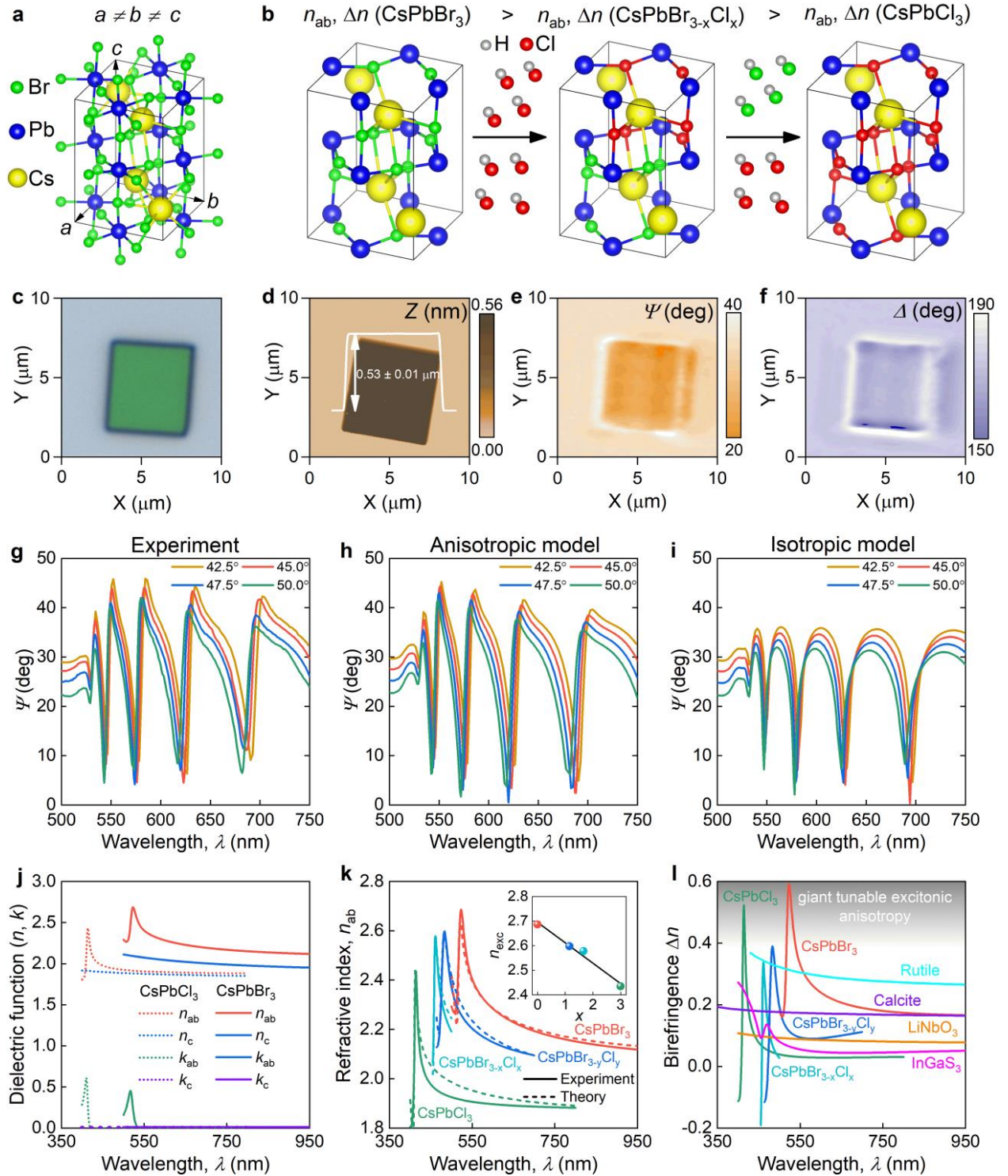

**Figure 1. Tunability of halide perovskites dielectric tensor. (a)** Orthorhombic crystal structure of $CsPbBr_3$. **(b)** Schematic illustration of chemical modification of halide perovskites. Representative **(c)** optical image, **(d)** atomic-force microscopy topography and cross-sectional profile, ellipsometric **(e)** $\Psi$ and **(f)** $\Delta$ maps, recorded at incident angle $\theta = 50°$ and $\lambda = 400$ nm, of $CsPbBr_{3-y}Cl_y$ microplate. **(g)** Experimental and theoretical ellipsometry spectra $\Psi$ of $CsPbBr_3$ based on **(h)** anisotropic and **(i)** isotropic model computations. **(j)** Anisotropic optical constants of $CsPbCl_3$ and $CsPbBr_3$. **(k)** Tunable refractive index of halide perovskite microplates for crystallographic *ab*-plane. The inset shows the dependence of the refractive index at exciton resonance $n_{exc}$ on stoichiometry number *x* of Cl atoms in $CsPbBr_{3-x}Cl_x$. **(l)** Tunable birefringence of halide perovskites in comparison with other anisotropic materials: rutile,[39] calcite,[38] $LiNbO_3$,[40] and $InGaS_3$.[37]

**Optical anisotropy in perovskite macrocrystals**

It is worth noting that anisotropic properties of halide perovskites are not limited to microplates and can be found in macrocrystals as well. For this purpose, we investigated $CsPbBr_3$ macroscopic samples (see the inset in **Figure 2c**), synthesized via anti-solvent vapor assisted crystallization (**Methods**).[41] The monocrystal nature of $CsPbBr_3$ sample was confirmed by X-ray diffraction (XRD) measurements. XRD spectrum, presented in **Supplementary Note 6**, which demonstrates four dominant diffraction peaks at 15.2°, 30.6°, 46.7°, and 63.9°, related to (101), (202), (303), and (404) plane of orthorhombic $CsPbBr_3$ perovskite with Pnma space group.

For anisotropy investigation, we recorded polarized reflection, plotted in Figure 1a. It immediately indicates that our samples are anisotropic because of clear reflectance sinusoidal behavior upon polarization. To get the qualitative description, we fitted experimental spectra (Figure 1a):

$$R(\lambda, \theta) = \left(\frac{n_c(\lambda) - 1}{n_c(\lambda) + 1}\right)^2 \cos^2(\theta) + \left(\frac{n_{ab}(\lambda) - 1}{n_{ab}(\lambda) + 1}\right)^2 \sin^2(\theta) \quad (1)$$

where $R(\lambda, \theta)$ are reflection coefficient at a given wavelength $\lambda$ and polarization angle $\theta$, $n_{ab}(\lambda)$ and $n_c(\lambda)$ are wavelength-dispersive refractive indices of CsPbBr$_3$. **Figure 2b** shows the resulting theoretical reflectance and **Figure 2c** demonstrates an exemplified fitting of polarized transmittance at given wavelength $\lambda$ = 900 nm. Similar to microplates, large monocrystal CsPbBr$_3$ exhibits higher refractive index along crystallographic ab-plane $n_{ab}$ compared to crystallographic c-axis $n_c$, as seen from **Figure 2d**. As a result, anisotropic properties of halide perovskites can be implemented not only in nanophotonics, but also in bulky optical devices such as waveplates or polarizers.

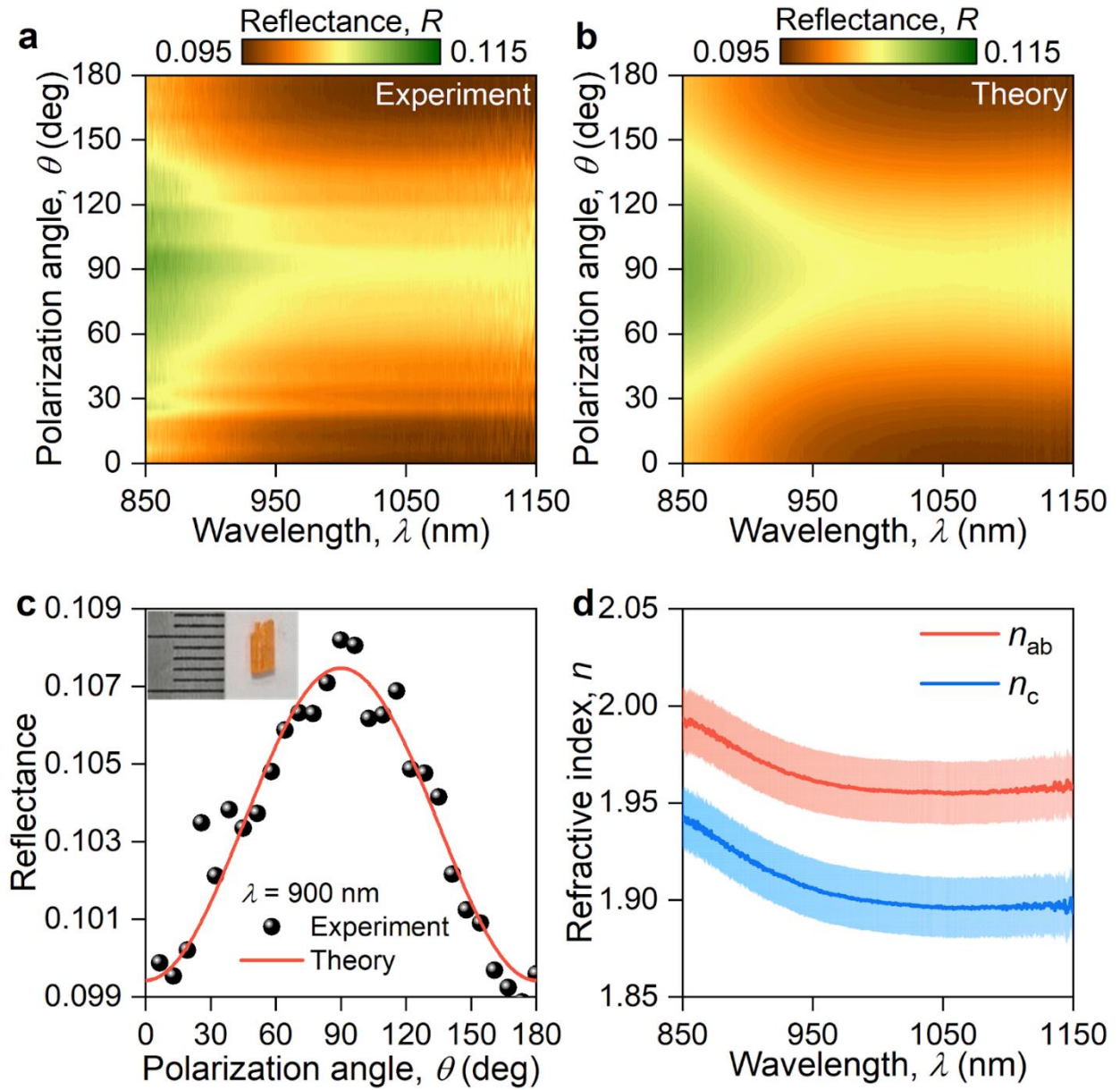

**Figure 2. Anisotropic response of CsPbBr₃ macrocrystal.** (**a**) Experimental and (**b**) theoretical polarized reflection of CsPbBr₃ single-crystal. (**c**) Representative reflection spectrum variation at λ = 900 nm with respect to polarization angle: θ = 0° (180°) and 90° correspond to polarization along crystallographic c-axis and ab-plane, respectively. The inset shows the photo of the investigated CsPbBr₃ crystal. (**d**) Anisotropic optical properties of CsPbBr₃ macrocrystal, obtained from fitting of panel (a). Error bars in the graph are taken from fitting errors.

**Anisotropic nanowaveguide**

Optical anisotropy is also critical for nanophotonic designs[42] where geometrical sizes move to subwavelength scales. In order to show the strong effect of anisotropy at nanoscale, we consider a CsPbBr₃ nanowaveguide (NW) with a diffraction grating on its top face. Depending on its period such grating allows for efficient angle-dependent coupling of the incident light into propagating modes of NW at a certain wavelength $\lambda_{in}$ (**Figure 3a**). Solution-synthesized monocrystalline CsPbBr₃ nanowires (protocol is available elsewhere[43]) on an ITO substrate were used as a base for such NWs. Structural analysis on the synthesized nanowires confirmed that their crystallographic axes are oriented perpendicularly to those in the above-mentioned CsPbBr₃ microplate (see **Supporting Note 7**), i.e. $n_c$ is oriented along the z-axis and $n_a = n_b$ are in x-y plane. Fs-laser projection lithography was further applied to imprint surface grating on the top face of the CsPbBr₃ nanowire[43]. An example of 20-um length nanowire with a rectangular cross-section (1 x 0.47 um²) containing laser-printed 260-nm period surface grating on the top face is shown in **Figure 3b**. Noteworthy, the laser patterning does not cause any detectable changes of either crystalinity or photoluminescence quantum yield of the nanowire (see **Supporting Note 8**).

We further analyzed the spectrum of the CsPbBr₃ NW modes that were excited by a collimated linearly polarized beam from a supercontinuum generator at different angles of incidence $\theta$ and out-coupled the NW from one of its facets (for details, see **Supporting Note 9**). This allows us to reconstruct the dispersion of the main waveguiding modes supported by the NW. The measurement results are summarized in **Figure 3c**, revealing several bright branches that can be attributed to certain modes supported by the CsPbBr$_3$ nanowire. Theoretical mode analysis for an infinitely long CsPbBr$_3$ NW with a rectangular-shaped cross-section (**Figure 3c,d**) was carried out to calculate the dispersion characteristics of the waveguiding modes. We considered the first-order diffraction condition $\Lambda(\lambda_{in}) = \lambda_{in}/[n_{eff}(\lambda_{in})-\sin\theta]$ and anisotropic refractive index of the CsPbBr$_3$ (**Figure 1j**). The calculation results for the strongest EH$_{00}$-, EH$_{10}$- and EH$_{20}$-modes are plotted in the **Figure 3c** as solid curves demonstrating perfect matching of calculated and the experimentally measured dispersion curves in the case of EH$_{10}$- and EH$_{20}$-modes. Such matching can not be achieved with any isotropic refractive index of the CsPbBr$_3$ (dashed curves in **Figure 3c**) indicating strong optical anisotropy of the NW. Noteworthy, the dispersion curve corresponding to the fundamental EH$_{00}$-mode is also marked in the **Figure 3c**. However, this mode can not be experimentally detected owing to weaker (with respect to EH$_{10}$- and EH$_{20}$-modes) excitation efficiency and the mismatch between its out-coupling directivity and the collection angle of the microscope objective (≈100°, NA=0.8) used in the experimental setup (**Figure 3d** and see for details **Supporting Notes 9 and 10**).

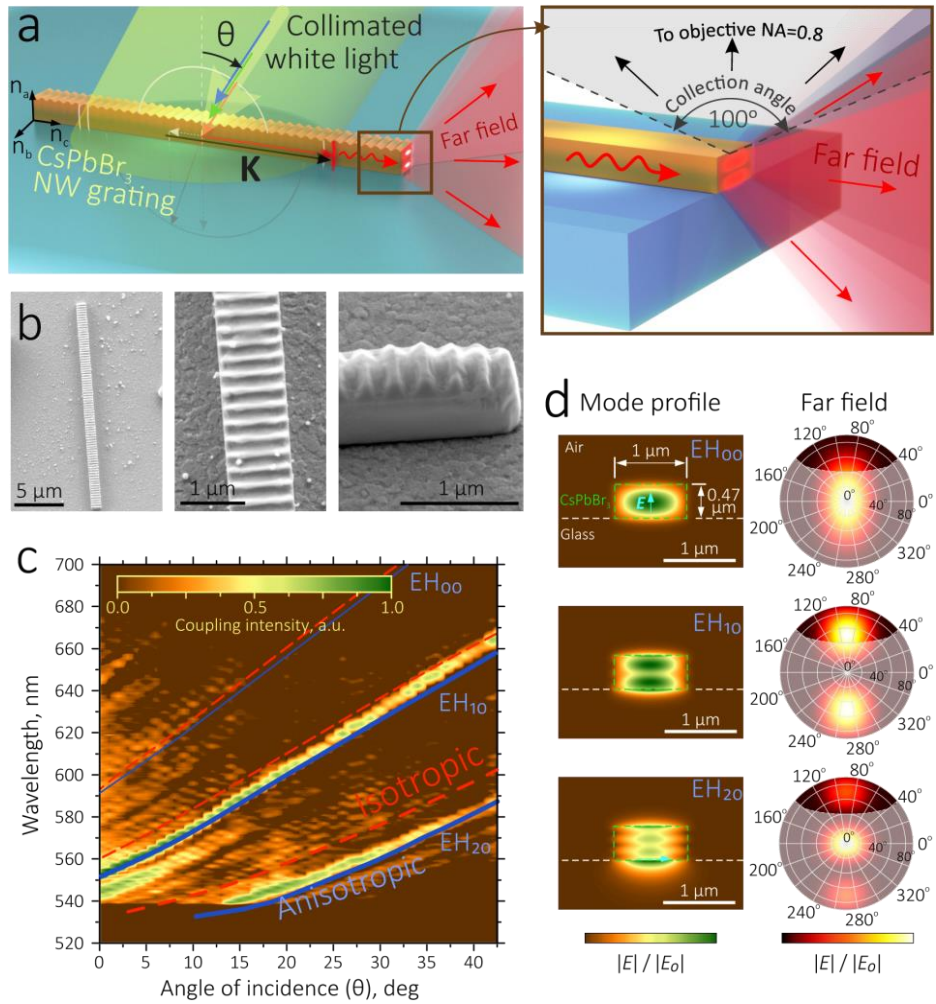

**Figure 3. Optical anisotropy in CsPbBr$_3$ nanowaveguide. (a)** Schematic illustration of the angled excitation of the CsPbBr$_3$ perovskite nanowaveguide as well as out-coupling of the modes. **(b)** SEM images of the perovskite nanowaveguide with a 260-nm period nanograting. **(c)** Measured and calculated dispersion law for the first three (EH$_{00}$, EH$_{10}$ and EH$_{20}$) waveguide modes supported by the perovskite nanowaveguide. Calculations were carried out considering both the anisotropic (solid curves) and isotropic (dashed curves) nanowaveguide. **(d)** Calculated intensity profiles and far-field outcoupling directivities for the considered waveguide modes.

## DISCUSSION

In summary, we measured the refractive index tensor of CsPbBr$_3$ for the first time and demonstrated its anisotropic optical response at all scales from nano to micro. This anisotropy originates from crystal low-symmetry and selective direction of exciton formation. Surprisingly, $CsPbBr_3$ exhibits the highest birefringence (up to 0.6) among non-van der Waals materials. Apart from giant anisotropy, this birefringence can be modified by chemical substitution of Br with Cl atoms. As a result, anisotropy can take any value from $CsPbBr_3$ to $CsPbCl_3$ dielectric tensor since chemical reaction results in intermediate state $CsPbBr_{3-x}Cl_x$ with continually tuned Br and Cl concentrations. Finally, halide perovskite can change it optical axis, resulting in in-plane and out-of-plane anisotropy, depending on the shape. Hence, halide perovskites greatly expand the library of anisotropic materials and present novel opportunities for tunable anisotropic nanophotonics.

## ACKNOWLEDGMENTS

S.V.M. and A.P.P. acknowledges the Priority 2030 Federal Academic Leadership Program. Ellipsometry characterization of anisotropy (G.A.E.) was supported by the Russian Science Foundation (grant № 22-29-01192).

## AUTHOR CONTRIBUTIONS

A.P.P., A.D.C., A.V.A., K.S.N., S.V.M., and V.S.V. suggested and directed the project. G.A.E., A.P.P., A.Y.Z., A.A.K., A.S.S., D.S., E.Z., G.T., performed the measurements and analyzed data. A.P.P., A.I., K.K., D.S. designed and fabricated the samples. I.K., A.M., and I.V. I. provided theoretical support. G.A.E., A.P.P., A.Y.Z., A.A.K., I.V.I., I.K., A.M., A.I., D.S., G.T., A.D.C., A.V.A., K.S.N., S.V.M., and V.S.V. contributed to the interpretation of the experimental results. G.A.E., A.P.P., S.V.M., wrote the original draft. G.A.E., A.P.P., A.I., I.V.I., A.D.C., A.V.A., K.S.N., S.V.M., and V.S.V. reviewed and edited the paper. All authors contributed to the discussions and commented on the paper.

## COMPETING INTERESTS

The authors declare no competing interests.

## METHODS

**Ellipsometry and microtransmittance measurements.** For ellipsometry measurements we used Accurion nanofilm_ep4 ellipsometer in rotating compensator mode with focus scan to ensure that the sample surface is in focus. Measurements were done in 1 nm step and at several incident angles (42.5°, 45°, 47.5°, 50°).

**Microtransmittance measurements.** Transmittance was recorded on Accurion nanofilm_ep4 ellipsometer at 90° position of input and output branch. First, we measured the camera signal, corresponding to sample, then reference and background signals from the same area. Transmittance was calculated as a ratio of sample to reference signals with subtracted background contribution.

**Reflectance measurements.** The reflectance spectra were obtained using a Hyperion 2000 microscope incorporated in Bruker Vertex 80 V spectrometer. For reference we used a gold mirror.

**Optical microscopy.** The images of halide perovskite microplates were captured by an optical microscope (Nikon LV150L) with a digital camera DS-Fi3.

**Atomic force microscopy.** AFM measurements were performed in air by a commercial AFM (NT-MDT N'tegra II, Moscow, Russia www.ntmdt-si.ru) The topography data was registered in semi-contact mode by Nova_PX acquisition software integrated with the AFM instrument using HA_NC cantilevers with tip-tapping resonant frequency around 145 kHz. The image processing and quantitative analysis were done using Gwyddion software (www.gwyddion.net).

**Synthesis of halide perovskite microwires and microplates.** Perovskite microwires on glass and ITO substrates were synthesized by using a protocol similar to the previously reported one[27]. 110 mg of $PbBr_2$ and 62 mg of CsBr were mixed and dissolved in 3 ml of anhydrous dimethyl sulfoxide (DMSO) inside a $N_2$-filled glove box. 2 μl of the prepared solution was drop-casted on ITO substrate at ambient conditions. Then, the substrate was sealed in a preheated up to 60 °C Petri dish containing 200 μl of liquid 2-propanol-water azeotropic mixture (85%). The droplet was dried in the presence of azeotropic vapor at 60 °C for 5 min. As a result ensembles of separate microwires were formed on the substrate.

Perovskite microplates on amorphous $Al_2O_3$ substrate with island morphology were synthesized by utilizing the same solution in which 20 μl of deionized water was added (Supplementary Figure 5). The protocol is the same to the above mentioned.

**Synthesis of halide perovskite crystal.**

Perovskite monocrystals were synthesized via anti-solvent vapor-assisted crystallization (AVC), described by Dmitry Dirin and co-authors.[41] Bromide precursors CsBr and $PbBr_2$ were dissolved in dipolar aprotic solvent with excess of Cs component. 56.82 mg of CsBr and 146.80 mg of $PbBr_2$ were dissolved in 1 ml of dimethyl sulfoxide (DMSO). The solution was prepared under argon atmosphere and had been stirred overnight at room temperature. The solution was filtered through PTFE 0.45μm syringe filters before the growth process, then 0.3 ml was added into the solution.

Antisolvent mixture was made by mixing ethanol and water in 1:6 volume ratio. The growth kit was set as a russian doll: perovskite precursor solution in small vial with no cap was placed inside bigger antisolvent vial. We observed that the growth rate and crystals quality depend on gas-liquid interface of solvent and antisolvent. The variety of vials size ratio were tested. The best result was achieved with 4 and 20 ml vials. The set up was placed on marble antivibration table at the room temperature (295K) until crystals reached the diameter of small vial. The crystal had plate shape, the dimension was 9 mm length, 2 mm with and 1 mm tall.

**X-ray diffraction.**

To confirm the monocrystal nature of the grown $CsPbBr_3$ samples, we performed the XRD analysis with Cu source. The XRD spectra of bulk monocrystals (Supplementary Note 3) show four dominant diffraction peaks at 15.2°, 30.6°, 46.7° and 63.9°, which can be related to 101, 202, 303, and 404 planes of orthorhombic $CsPbBr_3$ perovskite with Pnma space group.[44]

Investigation of structural properties for $CsPbBr_3$ MNCs. XRD patterns were recorded on Brucker D8 Discover X-ray diffractometer with Cu Kα radiation (1.54184 Å).

The XRD patterns of microwires (Supplementary Figure 8a) and microplates (Supplementary Figure 4a) ensembles were recorded in the θ–θ geometry on an XRD-7000 diffractometer (Shimadzu) equipped with a 2 kW rotating Cu anode tube, beam collimator and optical microscope. The X-ray beam was collimated to the 100 μm diameter spot and the measurements were carried out in the scan range 2θ = 10–60° with 0.002° precision. It was established microwires and microplates possess orthorhombic structure and crystallize in two different space groups Pbnm and Pnma, respectively.

**Transmission electron microscopy.** TEM, HRTEM, SAED and FFT images (Fig. S8b-d, Fig. S9b,c) were obtained on an FEI Tecnai F20 X-TWIN transmission electron microscope with a field emission gun. Sample measurements were carried out using the bright-field regime, acceleration tension set for 200 kV, images recorded using a Gatan Orius CCD camera. Perovskite crystals were synthesized on TEM grids according to the protocols employed for the production of microwires and microplates. The images confirm a monocrystalline structure of perovskite crystals (for details, see Supplementary Note 1).

**Fabrication and characterization of the anisotropic CsPbBr$_3$ nanowaveguide**

Direct femtosecond-laser (180-fs pulse duration, 515-nm wavelength) projection lithography was used to imprint the surface grating on the top face of the CsPbBr$_3$ nanowire. The output laser beam was first shaped into a flat-top stripe-shape beam using an aperture drilled in the Al foil. The generated intensity profile was then projected with magnification from the aperture plane to the microscope objective focal plane using 4*f*-optical system yielding in generation of 400-nm width 5-μm length laser stripe with uniform intensity distribution. Single-pulse single-trench laser patterning regime of the CsPbBr$_3$ nanowire with such stripe-shaped beam at constant scanning speed and pulse repetition rate was used to imprint the nanograting with the desired period. Additional details can be found elsewhere[44].

**Optical modeling.**

Intensity profiles of the certain waveguide modes supported by the $CsPbBr_3$ nanowire with a rectangular cross-section as well as out-coupling directivities of these modes were calculated using commercial Lumerical Mode Solutions software package.

## DATA AVAILABILITY

The datasets generated during and/or analyzed during the current study are available from the corresponding author on reasonable request.